# Memory of Quark Matter Card Game

J. Csörgő[1], Cs. Török[1] and T. Csörgő[2]

[1]Berze Science Club, Berze Middle School, H-3200 Gyöngyös, Kossuth u 33, Hungary

[2]MTA Wigner Research Center for Physics, Institute for Particle and Nuclear Physics

H- 1525 Budapest 114, POBox 49, Hungary

**Abstract and also a brief Introduction:**

The Top Physics Story for 2005, as selected by the American Institute of Physics [1], concerned the discovery of a nearly perfect fluid in high energy collisions of gold nuclei at Brookhaven National Laboratory's (BNL) Relativistic Heavy Ion Collider (RHIC) [2-5]. During 2010, RHIC scientists determined that the initial temperature of this perfect, liquid like soup of quarks and gluons reaches at least 4 trillion degrees Celsius, around 250 000 times the temperature at center of the Sun [6]. By 2012, this measurement became known as a scientific Guiness record, the hottest temperature ever made by humans [7]. The Hagedorn-temperature, the upper limiting temperature of 2 trillion degrees Celsius for the existence of protons, neutrons and other hadrons has been provenly exceeded, for the first time.

An educational aspect of this RHIC perfect fluid discovery is that the hottest known form of matter is not a gas, but behaves just like a fluid, that expands and flows much more perfectly than water or any other well known fluid. This aspect of the RHIC discovery can be introduced even to primary and secondary levels of physics education [8,9]: the solid to liquid to gas sequence of phase transitions now are known to be followed by a transition to a perfect, and liquid state of quarks at the largest temperatures made by humans.

The educational games described herein were invented by middle school students, members of a Science Club in Hungary. The games were invented for their entertainment, the educational applications in teaching high energy particle and nuclear physics to laypersons are quite unexpected but most welcomed. This manuscript is an updated version of a handout booklet, distributed by BNL's GUV Center starting from 2011. It includes an important contribution by Angela Melocoton, an administrator of BNL's Guests, Users and Visitors (GUV) Center, and describes in layperson's terms, how to play with quarks, antiquarks and leptons the "Memory of Quark Matter" style card games [9,10].

# MEMORY OF QUARK MATTER CARD GAME

**Number of players:** arbitrary.

**Object of the game:** to collect as many cards as possible by detecting particles as quickly as possible from the Quark Matter – the perfect fluid of quarks, by following its time evolution.

**The course of the game:** The players put the thoroughly shuffled pack to the middle then make a small area heap from them – with the particle sides of the cards, in a face up position. The stock of cards represents the formation of Quark Matter early in the game. Initially, neutrinos leave the plasma undetected, so the players pick out them first, but they do not get scores for it and can take neutrinos one by one. Then we detect the dilepton enhancement, generated by the enormous initial temperature. These dileptons are detected as they escape from the quark-gluon plasma without strong interactions. The players refer to this by searching for lepton–anti-lepton pairs (e⁻e⁺ or µ⁻µ⁺ pairs). Then we arrive to the phase of hadronization. Players form various baryons and mesons from the quarks and anti-quarks left in the middle, until all the quarks are used up.

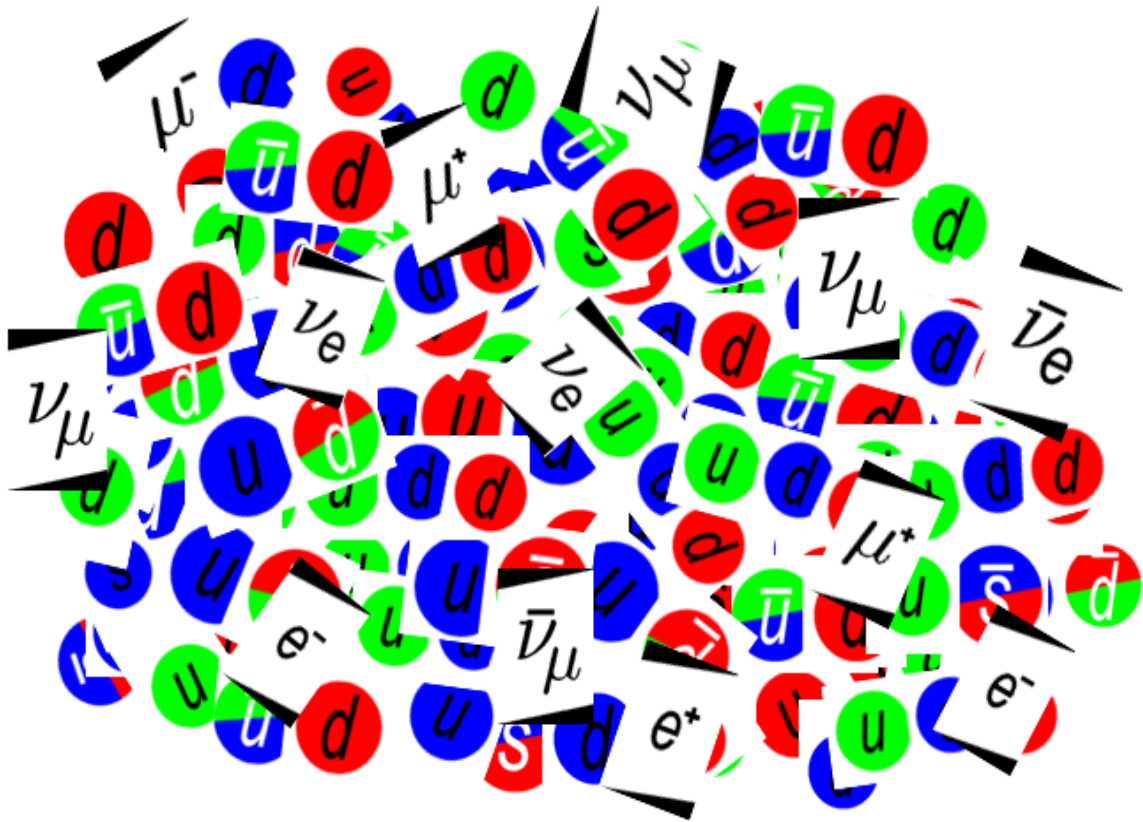

**Figure 1.** Illustration of the initial configuration for the card game Quark Matter.

**The Quark Matter card game can be played on various levels, even on a layperson's level.**

1) **On a beginner level,** players do not know the name of the hadrons that they form from quarks and anti-quarks. The only concern is to identify neutrinos, then lepton pairs, and then assure color neutrality when forming hadrons. Hadrons come either as mesons, or (anti)baryons. Mesons are created by a quark and anti-quark pair so that all the three colors are present (e.g. pairing blue quark with a red/green anti-quark). Baryons are fromed from three cards, representing a red, a green and a blue quark, while anti-baryons are formed from three antiquark cards, colored as red/green, green/blue and blue/red.

2) **On an intermediate level,** the goal is the get to know these names, so players can use a table of the particles to learn these names during the game. In this game mode, players follow each other clockwise, forming one hadron each time as the game proceeds, and players have to name the hadron that they have formed, otherwise they are passed by. The game has no winner at this level, as we mentioned the goal is to familiarize oneself with the name of the hadrons.

3) **On an advanced level**, the players know the name of the particles (or the name of many of them). In this case, there is no order between the players, and they can pull out lepton pairs then hadrons as quickly as possible, without waiting for one another. They have to collect the lepton pairs or the hadrons face up before them, separating the valid card combinations. When there is no more card in the middle, the players show their lepton pairs and hadrons one by one and tell their names. If it's correct then the player gets the point for that hadron. The winner is that player, who has collected the largest number of lepton pairs and hadrons.

4) **On the level of a layperson,** the players may start to put all cards face down on the table to play a "Memory" style game. The first player turns two cards face up. In the usual "Memory" style games, players can keep the cards if both are the same. However in a Quark Matter Memory game, the players can keep cards only if a valid particle pair (or triplet) is formed. Valid pairs come in two forms: either two black-and-white cards that represent lepton pairs, namely electron-positron, muon-antimuon or neutrino-anti-neutrino pairs, or hadrons. Hadrons come in two forms, mesons and baryions. Both can be formed from the colored cards. Mesons are formed from quark-antiquark cards. Baryons are formed from three quark cards. The quark cards have either a red, a green or a blue color, while the antiquarks are marked with two colors. Valid mesons and baryons too feature all the three (red, green, blue) colors. If the first two cards are both (anti)quarks (both represented with a red, a green or a blue colored card of if both are two-colored) and if their colors are different, the players can turn on a third card to try to form a baryon (or anti-baryon). Players can keep all the three cards if they are all quarks (or all anti-quarks) and all the three colors are represented. If the pairs or triplets of cards are not valid combinations, than players have to return them again with faces down. Players follow each other in an agreed (e.g. clock-wise) order untill all the cards are collected. The goal is to collect the largest number of cards.

**Thanks and credits are due Angela Melocoton** of Guests, Users and Visitors (GUV) Center, Brookhaven National Laboratory to for proposing this – "Memory" style – variation of the Quark Matter card game.

Quark Matter Card Games

Three more games, entitled `**ANTI!**', `**Cosmic Showers**' and `**Let's Detect!**' as well as some inspirational physics are described by the same authors in ref. [9]. Recently, a "Higgs Boson on Your Own" card game, based on the same Quark Matter Card Game deck of cards, was also developed and described in [10].